\documentclass[pra,aps,floatfix,showpacs,tightenlines,twocolumn]{revtex4}
\usepackage{float}
\usepackage{amsfonts}
\usepackage{makeidx}
\usepackage{graphicx}

\begin{document}

\title{Anyonic statistics with continuous variables}
\author{Jing Zhang$^{1,\dagger}$, Changde Xie$^{1}$, Kunchi Peng$^{1}$,
Peter van Loock$^{2}$}
\affiliation{$^{1}$State Key Laboratory of Quantum Optics and Quantum
Optics Devices, Institute of Opto-Electronics, Shanxi University,
Taiyuan 030006, P.R.China\\
$^{2}$Optical Quantum Information Theory Group,
Institute of Theoretical Physics I and Max-Planck Research Group,
Institute of Optics, Information and Photonics,
Universit\"{a}t Erlangen-N\"{u}rnberg, Staudtstr. 7/B2,
91058 Erlangen, Germany}

\begin{abstract}
We describe a continuous-variable scheme for simulating the
Kitaev lattice model and for detecting statistics of abelian anyons.
The corresponding quantum optical implementation is solely based upon
Gaussian resource states and Gaussian operations, hence allowing for
a highly efficient creation, manipulation, and detection of anyons.
This approach extends our understanding of the control and
application of anyons and it leads to the possibility for
experimental proof-of-principle demonstrations of anyonic statistics
using continuous-variable systems.
\end{abstract}

\maketitle

\section{Introduction}

It is well known from statistical physics that
in three spatial dimensions, only two kinds of particles exist,
bosons and fermions. A double exchange of any of these particles,
topologically equivalent to braiding one around the other,
will have no effect in three dimensions. In two spatial dimensions,
however, braiding one particle around the other is no longer
equivalent to doing nothing. Hence, the quantum mechanical
wave function can acquire any phase through the braiding operation,
leading to the fractional statistics of exotic quasiparticles,
so-called anyons \cite{one}.
Such anyons may be realized in model
systems that have highly entangled ground states with topological
order \cite{two,three,four}. The investigation of anyonic fractional
statistics is not only of fundamental interest.
It also led to the new field of topological quantum computation,
where the intrinsic robustness of the ground state Hamiltonian
against local perturbations offers naturally given fault tolerance
for quantum information processing \cite{five,six,seven,eight}.

The concept of anyons first emerged in
connection with the fractional quantum Hall effect \cite{two}, which
occurs in a two dimensional electron gas at low temperatures being
subject to a perpendicular magnetic field. The observation of
anyonic features in these systems requires cooling to very low
temperatures where only the ground and a few excited states are
populated. At the same time, anyons emerge as spatially localized
many-body states that can be created and transported through the
many-body system via local operations.
Although some signatures have been observed in the
fractional quantum Hall systems \cite{eleven}, a direct observation
of fractional statistics associated with anyon braiding has not been
achieved there. Nonetheless, various theoretical proposals exist
\cite{twelve}, including a spin-model scheme based upon laser manipulation
of cold atoms in an optical lattice \cite{thirteen}. However,
currently, this scheme appears to be beyond current experimental
capabilities.

A very simple model Hamiltonian from which one can obtain abelian anyons
was proposed by Kitaev for spin-$\frac{1}{2}$ (qubit) systems \cite{five}.
This surface code model involves suitable combinations of
four-body interactions. As such interactions are still hard to achieve
on the level of the ground-state Hamiltonian, a more practical approach
to simulating the surface code model would be based upon the creation
and manipulation of multi-party entangled graph states \cite{forteen}.
Instead of cooling the system to its ground state, the ground and excited
states are created dynamically from graph states.
By focussing on the underlying multi-party entangled states,
this approach provides a testbed for demonstrating anyonic properties
independent of the presence of a Hamiltonian,
in a simpler and more efficient way compared to the traditional approaches
based on complex solid-state systems
(though in order to obtain the full robustness needed for fault-tolerant
topological quantum computation, the background Hamiltonian defining
the protected code space appears to be crucial).

The graph-state-based protocol was
demonstrated experimentally with six-photon \cite{fifteen} and
four-photon graph states \cite{sixteen}. As reflected by these experiments,
both theoretical and experimental research of anyonic
fractional statistics was mainly concentrated on two-level or
spin-$\frac{1}{2}$ systems. Here, we shall consider
anyonic statistics in continuous-variable (CV) systems \cite{seventeen},
i.e., for quantized harmonic oscillators representing, in particular,
quantized optical modes. In order to apply the graph-state-based approach
\cite{forteen} to CV systems, we shall consider CV cluster and graph states
\cite{eighteen}. The corresponding optical Gaussian states can be efficiently and
unconditionally generated using off-line squeezed-state resources
and linear optics \cite{eighteen,twenty-five}.
Various four-mode CV cluster states have been experimentally realized already
\cite{ninteen,yukawa}. By adding a non-Gaussian
measurement to the toolbox of Gaussian operations, i.e., homodyne measurements
within the model of measurement-based quantum computation
\cite{Raussendorf}, universal quantum computation is, in principle,
possible using CV Gaussian cluster states \cite{twenty}.

Here, within the CV setting, we will show that the anyonic ground state
(or surface code state) can be created using Gaussian resource states
(off-line squeezed states)
and linear optics. Generating quasiparticle excitations and
implementing the braiding and fusion operations only requires
simple single-mode Gaussian operations; also detection and verification
of the quasiparticle statistics (observation of the fractional phase)
can be achieved via Gaussian operations including homodyne detection.

The plan of the paper is as follows. First, in
Sec.~\ref{cvcomp}, we will briefly introduce the formalism
and notations used in the paper for describing CV quantum logic and
computation. In the following, we discuss a generation scheme
for the CV anyonic ground state based on CV graph states
(Sec.~\ref{cvgraph}) and how to achieve the anyonic excitations
including the braiding operations (Sec.~\ref{cvanyoncreation}).
In Sec.~\ref{cvanyondetection}, we will describe a detection scheme
to verify the fractional phase of the anyons;
in this measurement scheme, the global phase acquired
by the excited state through the braiding operation is turned into a
phase-space translation resulting in different output stabilizer
operators (corresponding to different position-momentum linear combinations)
depending on whether a braiding loop has been applied or not.
Finally, before the conclusion in Sec.~\ref{conclusion},
we discuss in more detail potential optical implementations of our
CV-based anyon realization (Sec.~\ref{optical}).

\section{Continuous-variable computation}\label{cvcomp}

In order to consider quantum logic over continuous variables, the
qubit Pauli $X$ and $Z$ operators are generalized to the
Weyl-Heisenberg (WH) group \cite{twenty-one}, which is the group of
phase-space displacements. This is a Lie group with generators
$\hat{x}=(\hat{a}+\hat{a}^\dagger)/\sqrt{2}$ (quadrature-amplitude
or position) and $\hat{p}=i(\hat{a}-\hat{a}^\dagger)/\sqrt{2}$
(quadrature-phase or momentum) representing, for instance, a single
quantized mode (qumode) of the electromagnetic field. These
operators satisfy the canonical commutation relation
$[\hat{x},\hat{p}]=i$ (with $\hbar=1$). In analogy to the qubit
Pauli operators, the single-mode WH operators are defined as
$X(s)=\exp(-is\hat{p})$ and $Z(t)=\exp(it\hat{x})$ with $s,t\in
\mathbb{R}$. The WH operator $X(s)$ is a position-translation
operator, which acts on the computational basis of position
eigenstates $\{|x\rangle; x\in \mathbb{R}\}$ as
$X(s)|x\rangle=|x+s\rangle$; $Z(t)$ is a momentum-translation
operator, which acts on the momentum eigenstates as
$Z(t)|p\rangle=|p+t\rangle$. These operators are non-commutative and
obey the identity
\begin{eqnarray}
X(s)Z(t)=e^{-ist}Z(t)X(s).\label{com}
\end{eqnarray}
In the following, in analogy to the qubit case, we denote the WH
operators as Pauli operators. The Pauli operators for one mode can
be used to construct a set of Pauli operators
$\{X_{i}(s_{i}),Z_{i}(t_{i}); i=1,...,n\}$ for n-mode systems. This
set generates the Pauli group $\mathcal{C}_{1}$. The Clifford group
$\mathcal{C}_{2}$ is the group of transformations that preserve the
Pauli group $\mathcal{C}_{1} $ under conjugation; i.e., a gate
$\emph{U}$ is an element of the Clifford group if
$\emph{UR}\emph{U}^{\dagger}\in\mathcal{C}_{1}$ for every
$\emph{R}\in\mathcal{C}_{1}$. The Clifford group $\mathcal{C}_{2}$
for continuous variables corresponds to the (semidirect) product of
the Pauli group and the linear symplectic group of all one-mode and
two-mode squeezing transformations \cite{twenty-one}. The CV gate
for switching between the position and the momentum basis is given
by the Fourier transform operator
$F=\exp[i(\pi/4)(\hat{x}^{2}+\hat{p}^{2})]$, with
$F|x\rangle_{x}=|x\rangle_{p}$. This gate corresponds to a
generalization of the Hadamard gate for qubits. The phase gate
$P(\eta)=\exp[i(\eta/2)\hat{x}^{2}]$ is a CV squeezing operation, in
analogy to a Clifford phase gate for qubits \cite{twenty-two}. The
qubit controlled phase (sign) gate is generalized to a
controlled-$Z$, $C_{Z}=\exp(i\hat{x}_{1}\otimes\hat{x}_{2})$. This
gate provides the basic interaction for two modes 1 and 2; at the
same time, it describes a quadrature quantum nondemolition (QND)
interaction. The set $ \{X(s), F, P(\eta), C_{Z}; s,\eta \in
\mathbb{R}\}$ generates the Clifford group. Transformations within
the Clifford group do not form a universal set of gates for CV
quantum computation. However, Clifford group transformations
(corresponding to Gaussian transformations mapping Gaussian states
onto Gaussian states) together with any higher-order nonlinear
transformation acting on a single mode (corresponding to a
non-Gaussian transformation with a Hamiltonian at least cubic in the
mode operators) form a universal set of gates
\cite{twenty-one,lloyd}. As we will show in the following, all
operations needed to demonstrate the anyonic fractional statistics
in CV systems are entirely based upon Clifford group (Gaussian)
transformations and homodyne measurements (Gaussian operations).

%
\begin{figure}
\centerline{
\includegraphics[width=3in]{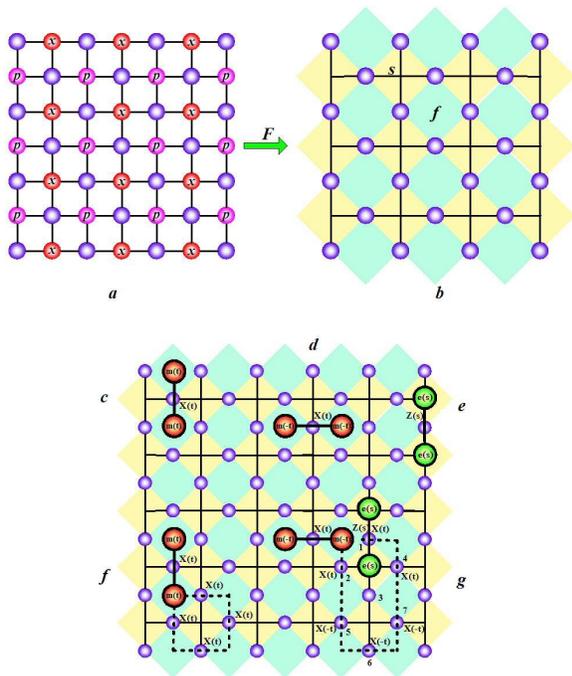}
} \vspace{0.1in}
\caption{(Color online). \textbf{Surface code model and anyonic
braiding with continuous variables}. \textbf{\emph{a}}, The
measurement pattern to prepare the anyonic ground state
$|\psi\rangle$ from a two-dimensional CV cluster state, where $x$
and $p$ denote a single-mode measurement in the position and
momentum basis, respectively, $F$ is the Fourier transform for all
remaining modes. \textbf{\emph{b}},The CV surface code model
consists of the star operators $A_{s}(\xi)$ and plaquette operators
$B_{f}(\eta)$. \textbf{\emph{c}}, Application of an $X(t)$
position-translation on a single mode in the horizontal edge yields
a pair of $m(t)$-type anyons placed at the neighboring plaquettes.
\textbf{\emph{d}}, Application of $X(t)$ on a single mode in the
vertical edge yields a pair of $m(-t)$-type anyons.
\textbf{\emph{e}}, Application of $Z(s)$ on a single mode yields a
pair of $e(s)$-type anyons at the neighboring stars.
\textbf{\emph{f}}, Application of $X(t)$ on the modes along a string
would create $m$-type anyons, where two $m$-type anyons are created
on the same plaquettes. These two anyons annihilate each other
according to the fusion rule $m(t)\times m(-t)=1$. Thus, the two
single strings are glued together with two anyons at the endpoints.
When a part of a string forms a loop, it cancels (dashed).
\textbf{\emph{g}}, An $m$-type anyon follows a closed loop around an
$e$-type anyon.
 \label{Fig1} }

\end{figure}

\section{Continuous-variable graph and surface code states}\label{cvgraph}

A graph quantum state is described by a mathematical graph, i.e., a
set of vertices connected by edges \cite{twenty-three}. A vertex
represents a physical system, e.g. a qubit (discrete two-level
system in a 2-dimensional Hilbert space) or a CV qumode (system with
a continuous spectrum in an infinite-dimensional Hilbert space). An
edge between two vertices represents the physical interaction
between the corresponding systems. The preparation procedure of CV
cluster states \cite{eighteen} is analogous to that for qubit
cluster states: first, prepare each mode (or graph vertex) in a
phase-squeezed state, approximating a zero-momentum eigenstate (the
analogue of the Pauli-$X$ `+1' eigenstate); then apply a QND
interaction [the gate $C_{Z}=\exp(i\hat{x}_{j}\otimes\hat{x}_{k})$]
to each pair of modes $(j,k)$ linked by an edge in the graph. Note
that all $C_{Z}$ gates commute. The resulting CV cluster/graph state
(and, more generally, any CV cluster-type state \cite{twenty-five})
satisfies, in the limit of infinite squeezing \cite{eighteen},
\begin{eqnarray}
\hat g_{a}=(\hat{p}_{a}-\sum_{b\in
N_{a}}\hat{x}_{b})\rightarrow 0, \quad\forall a\in G,
\end{eqnarray}
where the modes $a\in G$ correspond
to the vertices of the graph of $n$ modes and the modes $b\in
N_{a}$ are the nearest neighbors of mode $a$.
More precisely, this relation describes a
simultaneous zero-eigenstate of the $n$ position-momentum linear
combination operators $\hat g_{a}$.

The stabilizer operators $G_{a}(\xi)=\exp(-i \xi \hat
g_{a})=X_{a}(\xi)\prod_{b\in N_{a}}Z_{b}(\xi)$ with $\xi\in
\mathbb{R}$ for CV cluster states are analogous to the $n$
independent stabilizers $G_{a}=X_{a}\prod_{b\in N_{a}}Z_{b}$ for
qubit cluster states \cite{twenty-three}. Starting from a 2D CV
cluster state, single-mode measurements of every second mode in
either the position or the momentum basis, with a subsequent Fourier
transform $F$ for all the remaining modes, yields a new graph state
$|\psi\rangle$. The measurement pattern is illustrated in Fig.1a. An
$x$-measurement removes the measured mode from the cluster and
breaks all connections between this mode and the rest of the
cluster, including momentum-displacements of the nearest modes with
the measured result. Conversely, the effect of a $p$-measurement is
to remove the measured mode, but to connect the neighboring modes at
the same time. These measurements transform the set of correlations
in terms of position-momentum linear combination operators from
$\{(\hat{p}_{a}-\sum_{b\in N_{a}}\hat{x}_{b})\rightarrow 0\}$ to
$\{\hat
a_{s}=(\hat{p}_{s,1}+\hat{p}_{s,2}+\hat{p}_{s,3}+\hat{p}_{s,4})
\rightarrow 0, \hat b_{f}=
(\hat{x}_{f,1}-\hat{x}_{f,2}+\hat{x}_{f,3}-\hat{x}_{f,4})\rightarrow
0\}$ for the remaining modes, where $s$ and $f$ label stars and
plaquettes (faces), respectively, and the indices 1,...,4 of the
position and momentum operators denote those modes located at a
common star or at the boundary of a common plaquette. The
stabilizers $\{A_{s}(\xi)=\exp(-i \xi \hat a_{s})=\Pi_{j\in
star(s)}X_{s,j}(\xi),B_{f}(\eta)=\exp(-i \eta \hat b_{f})=\Pi_{j\in
boundary(f)}Z_{f,j}((-1)^{j}\eta)\}$, with $\xi,\eta\in \mathbb{R}$,
for the new CV state $|\psi\rangle$ are exactly analogous to those
of the first Kitaev model \cite{five} for a two-dimensional spin
lattice. These stabilizers, the star operators $A_{s}(\xi)$ and
plaquette operators $B_{f}(\eta)$, all commute. So the new CV state
$|\psi\rangle$ corresponds to the anyonic ground state with
$A_{s}(\xi)|\psi\rangle=|\psi\rangle$ and
$B_{f}(\eta)|\psi\rangle=|\psi\rangle$, $\forall \xi,\eta\in
\mathbb{R}$, for all stars $s$ and plaquettes $f$, in the limit of
infinite squeezing. After defining the CV anyonic ground state and
describing how to create an approximate version of this state
(asymptotically perfect in the limit of infinite squeezing) from CV
cluster states, let us now consider anyonic excitations from the
ground state and the basic braiding operations.

\section{Anyon creation and braiding}\label{cvanyoncreation}

Starting from the ground state
one can excite pairs of anyons connected by a string using
single-mode operations. Unlike the Pauli group for qubits, the
WH group is a continuous group, and therefore
quasiparticle excitation is continuous. More specifically, by
applying the position-translation operator $X(t)$ to some mode of
the lattice, a pair of so-called $m$-type anyons in the state
$|m((-1)^{d}t)\rangle=X(t)|\psi\rangle$ ($d\in\{1,2\}$, $d=1$ means
that the relevant mode lies on the vertical edges, whereas $d=2$ refers to the
horizontal edges) is created on the two neighboring plaquettes (Fig.1c,d).
An $e$-pair of anyons, given by
$|e(s)\rangle=Z(s)|\psi\rangle$, is obtained on two neighboring stars by a
momentum-translation operation $Z(s)$ (Fig.1e).

Similar to the qubit case, the excited states differ from the ground states
in their stabilizer operators. In the CV case, this means that some of the
position-momentum linear combination operators $\hat a_{s}$ and
$\hat b_{f}$ [from the ground-state stabilizers
$A_{s}(\xi)=\exp(-i \xi \hat a_{s})$ and
$B_{f}(\eta)=\exp(-i \eta \hat b_{f})$] no longer satisfy
the ground-state conditions $\hat a_{s} = 0$ and $\hat b_{f} = 0$
(in the limit of infinite squeezing), but rather become
$\hat a_{s} = t$ for $Z_{j\in
star(s)}(t)|\psi\rangle$ and $\hat b_{f} = \pm s$ for
$X_{j\in
boundary(f)}(s)|\psi\rangle$. In other words, the anyonic excitations
become manifest in ``violations'' of the stabilizer (``nullifier'')
conditions for $\hat a_{s}$ in the presence of $e$-particles and for
$\hat b_{f}$ in the presence of $m$-particles.

The fusion rules
[$e(s)\times e(t)=e(s+t), m(s)\times m(t)=m(s+t),e(0)=m(0)=1,
1\times e(s)=e(s), 1\times m(s)=m(s)$] describe the outcome from
combining two anyons. Thereby, excited states can be obtained from
the CV ground state $|\psi\rangle$ by applying open string
operators, which create quasiparticles at their endpoints (Fig.1f).
The closed $X$ ($Z$) strings surround the corresponding stars
(plaquettes), which are a product of the surrounded star (plaquette)
operators (Fig.1f). Thus, for any closed $X$ ($Z$) string, the system
will remain in the ground state.

The anyonic character of the excited states
is now revealed through a non-trivial phase factor
acquired by the wavefunction of the lattice system after braiding
anyons, i.e., after moving $m$ around $e$ (Fig.1g) or vice versa.
Consider the initial state
$|\Psi_{ini}\rangle=Z_{1}(s)|\psi\rangle=|e(s)\rangle$. If an anyon
of type $m$ is assumed to be at a neighboring plaquette, it can be
moved around $e$ along the path generated by successive applications
of $X(t)$ on the four modes of the star. The final state
is
\begin{eqnarray}
|\Psi_{fin}\rangle&=&X_{1}(t)X_{2}(t)X_{3}(t)X_{4}(t)|\Psi_{ini}\rangle
\nonumber \\
&=&e^{-ist}Z_{1}(s)[X_{1}(t)X_{2}(t)X_{3}(t)X_{4}(t)|\psi\rangle]
\nonumber \\&=&e^{-ist}|\Psi_{ini}\rangle,\label{phase}
\end{eqnarray}
using Eq.(\ref{com}).
The extra factor $\exp(-ist)$ is the topological phase factor, which reveals the
presence of the enclosed anyons (it would not be obtained, if the initial
state had been the unexcited state, $|\Psi_{ini}\rangle=|\psi\rangle$).
Note that here in the CV setting,
the quantum state can acquire any phase, in contrast to the discrete value
for anyons in qubit systems with the analogous braiding operation
(where $X Z = - Z X$ for qubits).

The corresponding phase factors, changing the
phase of the wavefunction, constitute an abelian (commutative)
representation of the braid group.
It is important to understand that, in general,
a relation as described by Eq.(\ref{phase}) can be obtained
for different paths of the loop. For instance, the loop
1-2-5-6-7-4 of $m$-type anyons (Fig.1g), which corresponds to the
application of
$X_{1}(t)X_{2}(t)X_{5}(-t)X_{6}(-t)X_{7}(-t)X_{4}(t)=
[X_{1}(t)X_{2}(t)X_{3}(t)X_{4}(t)][X_{3}(-t)X_{5}(-t)X_{6}(-t)X_{7}(-t)]$,
involving two star operators, gives the same result for the
topological phase as in Eq.(\ref{phase}).
In fact, this equivalence reflects the topological character of the extra
phase factor and the potential robustness of the corresponding state
transformation, as it is exploited in topological approaches
to fault-tolerant quantum computation.

\section{Anyon detection}\label{cvanyondetection}

In order to detect the most significant feature of anyons,
their non-trivial statistical phase obtainable through braiding,
we may employ an ``interference
measurement'' in phase space, which enables us to detect the
otherwise invisible global phase factor of Eq.(\ref{phase}).
The idea is to turn the global phase acquired by the excited state
into a displacement in phase space that would not occur without
braiding operation.

First, we generate a superposition
of the ground state $|\psi\rangle$ and the continuously-excited
state of the anyon $e$ via the squeezing operation
$P(s)=\exp[i(s/2)\hat{x}^{2}]$ (the CV quadratic phase gate) instead of the
momentum-translation operation $Z(s)=\exp(is\hat{x})$. Then we
perform a closed loop of anyon $m$ around $e$. Finally,
application of the inverse squeezing operation
$P(-s)=\exp[-i(s/2)\hat{x}^{2}]$ makes the phase difference visible,
leading to the state $e^{-it^{2}s/2}|e(st)\rangle$ according to
\begin{eqnarray}
e^{-i(s/2)\hat{x}^{2}}e^{-it\hat{p}}e^{i(s/2)\hat{x}^{2}}=
e^{-it^{2}s/2}e^{ist\hat{x}}e^{-it\hat{p}}.\label{phase2}
\end{eqnarray}
This compares to the state
$e^{-i(s/2)\hat{x}^{2}}e^{i(s/2)\hat{x}^{2}}|\psi\rangle=|\psi\rangle$
without a closed loop of anyon $m$ around $e$.
In other words, the effect of the braiding operation
can be measured via the different stabilizer operators or,
more precisely, the different position-momentum linear combination
operators of the output states: for the relevant star operator $s_0$,
we have $\hat a_{s_0}=st$ with braiding and $\hat a_{s_0}=0$
without braiding.

Note that this verification scheme is only complete,
provided that, in addition to detecting the change of the stabilizer
(nullifier) operator $\hat a_{s_0}$, a sufficient set of extra
nullifiers $\hat a_{s}$ and $\hat b_{f}$ is measured
in order to confirm the full inseparability of the output states
(the ground state and the momentum-displaced ground state) \cite{PvL}.
Otherwise, even a fully separable product state of zero-momentum
eigenstates used as ``ground state'' could produce the same
braiding statistics, as it would satisfy $\hat a_{s_0}=0$ without braiding
and $\hat a_{s_0}=st$ with a ``closed braiding loop''.
Such a trivial ground state, however, would obviously not
simulate Kitaev's surface code model; in particular, apart from
not satisfying the complete set of ground-state correlations, it would not
allow for a verification of excited $m$-particles or the ``conjugate''
braiding operation of moving $e$ around $m$.

%
\begin{figure}
\centerline{
\includegraphics[width=3in]{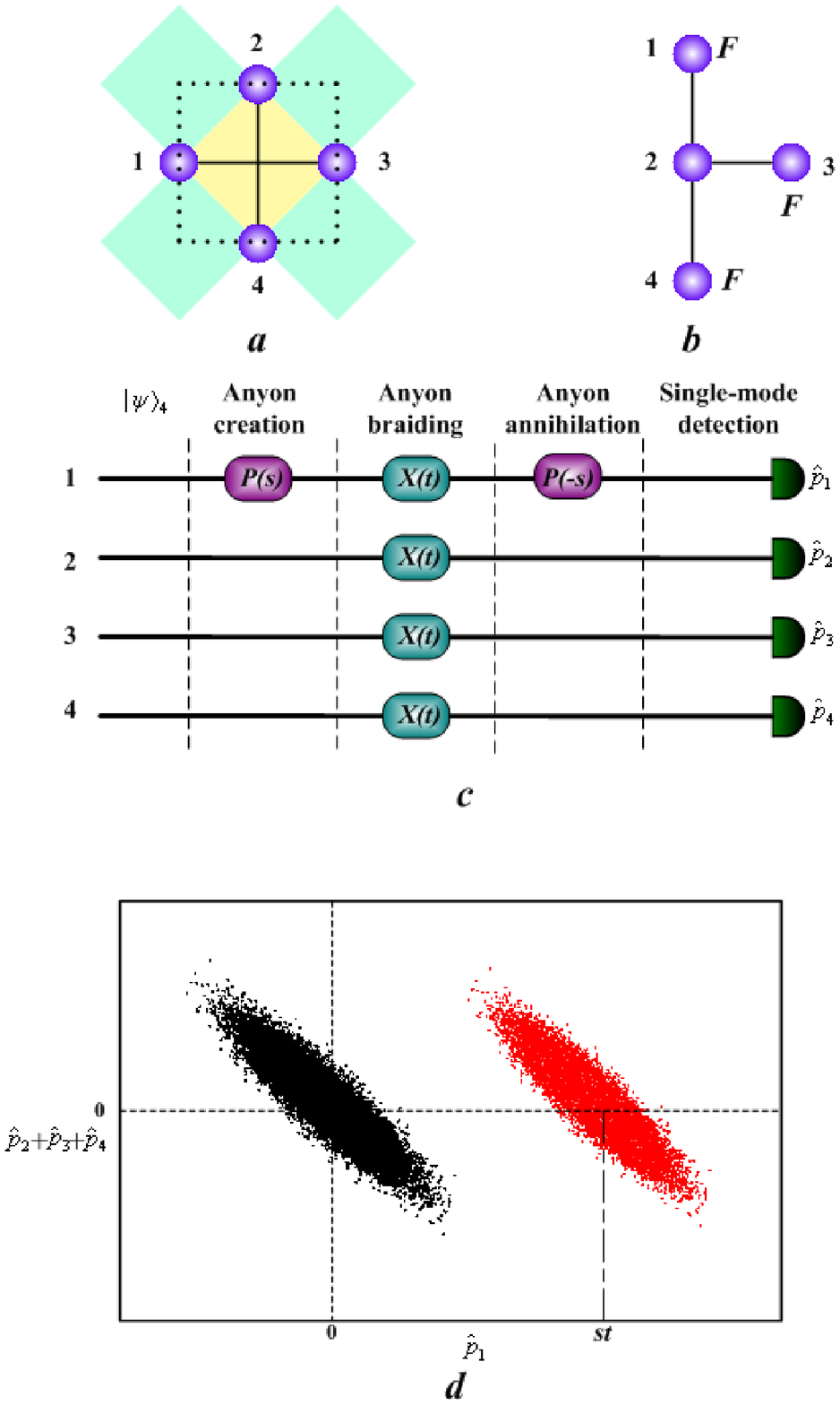}
} \vspace{0.1in}
\caption{(Color online). \textbf{Surface code state with four modes
used for demonstration of anyonic statistics.} \textbf{\emph{a}},
The anyonic ground state $|\psi\rangle_{4}$, forming one star and
four plaquettes, is given by a four-mode GHZ state with the modes
labelled 1 to 4. \textbf{\emph{b}}, The four-mode cluster state
which after Fourier transforms on modes 1,3, and 4, is equivalent to
the anyonic ground state $|\psi\rangle_{4}$. \textbf{\emph{c}}, The
quantum circuit for creation, braiding, and annihilation of the
anyons, and for the final state detection. \textbf{\emph{d}}, The
correlation diagram of the momentum linear combination operator
$\hat{p}_{1}+\hat{p}_{2}+\hat{p}_{3}+\hat{p}_{4}$ with (red) and
without a closed loop of anyon $m$ around $e$ (black) in the case of
finite squeezing.
 \label{Fig2} }

\end{figure}

\section{Optical implementation}\label{optical}

In a real experiment, because of finite squeezing,
it is impossible to create perfect
correlations in terms of position-momentum linear combination operators
(the stabilizers or, better, nullifiers) for the anyonic ground state
$|\psi\rangle$.
Nonetheless, we may still implement and detect anyonic
fractional statistics by interpreting the presence
of small fluctuations around the ground-state
stabilizer conditions due to finite squeezing, without
the existence of any first-moment phase-space translations,
as the ground state (Fig. 2d);
sufficiently large first-moment position (or momentum) translations,
still maintaining the second-moment multi-mode position-momentum
correlations, would then correspond to the excited states.

For verification, the amplitude (phase) quadrature,
i.e., the position (momentum) of each mode is measured via
homodyne detection and every detector output is sampled
within a time interval to yield a number of measured quadrature
values. Using the correlation diagram \cite{twenty-four} containing
a sufficient number of measured quadrature values, one can determine
first and second moments of the position-momentum linear combination
operators for the anyonic ground and excited states. Thus, the fully
inseparable ground states can be verified and discriminated from
the (for large squeezing near-orthogonal) excited (displaced ground)
states \cite{PvL}; at the same time enabling one to confirm
the braiding-dependent state transition according to
Eq.(\ref{phase2})(see Fig.2d). For finite, even small squeezing values,
it is still possible to discriminate the ground and excited states,
provided the amplitude (or phase) translations are sufficiently large.
Higher squeezing values improve the
sensitivity for detecting the amplitude (or phase) translations of
the excited states. Note that even for zero-squeezing resources, such a state
discrimination can be almost perfect (for large amplitudes/phases);
in this case, more subtle is the verification of the full inseparability
\cite{PvL} of both the ground and excited states, as the multi-party
entanglement in the approximate CV surface-code model would decrease
in this case (but it would not entirely vanish, because of the intrinsic
squeezing of the $C_{Z}$ gates for CV graph-state generation).

In the Kitaev lattice model,
the quasiparticles are particularly well localized: an $e$-particle
is on a single star and an $m$-particle is on a single plaquette. Based
on this perfect localization, the proposal for implementing the anyonic
model via the simple surface code has been demonstrated experimentally
with four-photon \cite{sixteen} and six-photon \cite{fifteen} graph
states. The four-photon experiment relies upon only four photonic qubits
entangled in a four-party GHZ state. This simple system is
sufficient for demonstrating a small Kitaev lattice (with one star
and four plaquettes) and a non-trivial braiding operation,
because the remaining eight qubits of the total 12-qubit system
factor out from the four-qubit GHZ state in a fully separable product state
\cite{sixteen}. In the CV regime,
we propose to pursue an analogous strategy.
In fact, four-mode entangled CV GHZ and cluster states
have been demonstrated already experimentally \cite{ninteen,yukawa}.
In a small-scale surface-code experiment, it is crucial
to verify the full inseparability of the GHZ-type ground state
as well as that of the excited state, i.e., a displaced GHZ state.
These simple systems may then also be used to
implement and detect minimal braiding loops (Fig. 2a,b,c) with four-mode
entangled GHZ states. The complete verification requires detecting
the position-momentum linear combinations corresponding to the
stabilizer set
\begin{eqnarray}
&&X_1(t) X_2(t) X_3(t) X_4(t),\nonumber\\
&&Z_1(t) Z_2^{-1}(t),
Z_2(t) Z_3^{-1}(t),Z_3(t) Z_4^{-1}(t),
\end{eqnarray}
including first and second moments \cite{PvL}.

However, for the simple four-mode GHZ states,
there is no way to show the robustness of the
topological braiding operation when the anyons are moved along different
paths. In order to demonstrate such robustness of the anyon operations
against different braiding paths, more modes would be needed, for instance,
nine-mode states in analogy to the nine-qubit states of Ref.\cite{forteen}.
The corresponding CV nine-mode ground state
$|\psi\rangle_{9}$ is locally equivalent to a CV nine-mode graph state,
which can be efficiently generated via nine off-line single-mode squeezed
states and arrays of beam splitters
\cite{twenty-five} instead of the projective measurements on a
prepared 2D CV cluster state as described previously \cite{twenty-six}.
In these larger lattices, there are several
different loops for the quasiparticles. The anyonic statistics, however,
would only depend on the topological character of the loop.

\section{Conclusion}\label{conclusion}

In summary, we have proposed a protocol to create, manipulate, and
detect anyonic states, including demonstration of their fusion rules
and their fractional statistics, in the regime of continuous
quantum variables. In our continuous-variable approach to
Kitaev's surface code model,
the main features of anyons can be obtained
via simple Gaussian operations starting from the creation
of suitable continuous-variable graph states.
With regard to a potential
optical realization, this means that all ingredients
for a full verification can be implemented in a very efficient way:
graph-state generation only requires off-line squeezing
and linear optics; the surface-code ground states are then obtained either
through homodyne measurements or via local phase rotations;
anyon creation and braiding-dependent state transitions
can be also detected through Gaussian operations, i.e.,
online squeezing transformations (which could be shifted off-line
\cite{filip}) and homodyne detections (including first and second
moments of quadrature combinations in order to verify
the fully inseparable ground and excited states and to discriminate
them against each other).

This work is a first step towards further theoretical investigations of
topological quantum computation over continuous variables
and experimental proof-of-principle demonstrations of anyonic, topological
features via relatively simple, stable optical systems,
as opposed to the more traditional approaches within condensed matter physics.

$^{\dagger} $Corresponding author's email address:
jzhang74@yahoo.com, jzhang74@sxu.edu.cn

\acknowledgments

This research was supported in part by NSFC for Distinguished Young
Scholars (Grant No. 10725416), National Fundamental Research Program
(Grant No. 2006CB921101), NSFC (Grant No. 60678029), NSFC Project
for Excellent Research Team (Grant No. 60821004), the PCSIRT (Grant
No. IRT0516), the TYMIT and TSTIT of Shanxi. PvL acknowledges
support from the Emmy Noether programme of the DFG in Germany.


\begin{thebibliography}{99}
\bibitem{twenty-six} a CV graph state as large as a nine-mode state
or even larger may also be created directly through a suitable nonlinear
interaction (without the need for complicated multi-mode interferometers)
\cite{menicucci}. Although this approach is particularly promising
for simulating larger Kitaev lattices based on larger graph states,
one complication appears to be how to address the different frequency
modes of the lattice for quasiparticle creation and braiding.
\end{thebibliography}
\end{document}